\documentclass[journal,twoside]{IEEEtran}

\usepackage{epsfig}
\usepackage{cite}
\usepackage{amsmath}
\usepackage{subfigure}

\newcommand{\bq}{\begin{subequations}}
\newcommand{\eq}{\end{subequations}}
\newcommand{\bqq}{\begin{subeqnarray}}
\newcommand{\eqq}{\end{subeqnarray}}
\newcommand{\beq}{\begin{eqnarray}}
\newcommand{\eeq}{\end{eqnarray}}
\newcommand{\beqq}{\begin{eqnarray*}}
\newcommand{\eeqq}{\end{eqnarray*}}
\newcommand{\be}{\begin{equation}\displaystyle}
\newcommand{\ee}{\end{equation}}

\begin{document}

\title{A Novel Boundary Element Method Using Surface Conductive Absorbers for Full-Wave Analysis of 3-D Nanophotonics}

\author{Lei Zhang, {\em Student Member, IEEE}, Jung Hoon Lee, Ardavan Oskooi, Amit Hochman, {\em Member, IEEE}, \\ Jacob K. White, {\em fellow, IEEE} and Steven G. Johnson
\thanks{Manuscript received \today.}
\thanks{L. Zhang, A. Hochman and J. K. White are with the Department of Electrical Engineering and Computer Science, Massachusetts Institute of Technology, Cambridge, MA, 02139, USA. (Email: zhangl@mit.edu; hochman@mit.edu; white@mit.edu)}
\thanks{J. H. Lee is with Applied Computer Science and Mathematics, Merck \& Co., Inc., Rahway, NJ, 07065, USA (Email: jung\_hoon\_lee@merck.com)}
\thanks{A. Oskooi is with the Center for Materials Science and Engineering, Massachusetts Institute of Technology, Cambridge, MA, 02139, USA. (Email: ardfar@mit.edu)}
\thanks{S. G. Johnson is with the Department of Mathematics, Massachusetts Institute of Technology, Cambridge, MA, 02139, USA. (Email: stevenj@math.mit.edu)}
}
\markboth{}
{Zhang \lowercase{{\em et al.}}: A Novel BEM Using Surface Conductive Absorbers for Full-Wave Analysis of 3-D Nanophotonics}

\maketitle

\begin{abstract}
Fast surface integral equation (SIE) solvers seem to be ideal approaches for simulating 3-D nanophotonic devices, as these devices generate fields both in an interior channel and in the infinite exterior domain. However, many devices of interest, such as optical couplers, have channels that can not be terminated without generating reflections.  Generating absorbers for these channels is a new problem for SIE methods, as the methods were initially developed for problems with finite surfaces. In this paper we show that the obvious approach
for eliminating reflections, making the channel mildly conductive outside the domain of interest, is inaccurate. We describe a new method, in which the absorber has a gradually increasing surface conductivity; such an absorber can be easily incorporated in fast integral equation solvers. Numerical experiments from a surface-conductivity modified FFT-accelerated PMCHW-based solver are correlated with analytic results, demonstrating that this new method is orders of magnitude more effective than a volume absorber, and that the smoothness of the surface conductivity function determines the performance of the absorber. In particular, we show that the magnitude of the transition reflection is proportional to $1/L^{2d+2}$, where $L$ is the absorber length and $d$ is the order of the differentiability of the surface conductivity function.

\end{abstract}


\section{Introduction}\label{sec:introduction}
\begin{figure}[b]
\begin{center}
\epsfclipon
\scalebox{0.8} {\rotatebox{270}{\epsffile{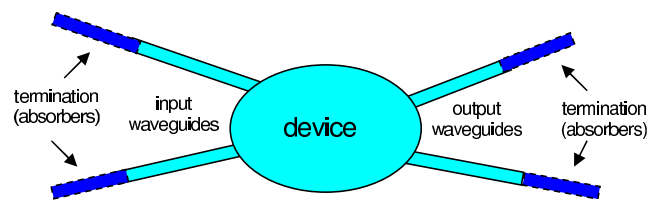}}}
\caption{Schematic diagram of a photonic device with input and output waveguide channels, which must be truncated in a surface integral equation method.}
\label{fig:device}
\end{center}
\end{figure}

In this paper we describe an absorber technique for terminating optical waveguides with the surface integral equation (SIE) method, which otherwise has difficulties with waveguides and surfaces extending to infinity. In order to attenuate waves reflected from truncated waveguides, we append a region with surface absorption to the terminations, as diagrammed in Fig.~\ref{fig:device}. The transition between the non-absorbing and absorbing regions will generate reflections that can be minimized by making the transition as smooth as possible. We show how this smoothness can be achieved with the surface absorber by smoothly changing integral-equation boundary conditions. Numerical experiments demonstrate that the reflections of our method are orders of magnitude smaller than those of straightforward approaches, for instance, adding a volume absorptivity to waveguide interior. In addition, we show the asymptotic power-law behavior of transition reflections as a function of the length of the surface absorber and demonstrate that the power law is determined by the smoothness of the transition \cite{oskooi08the}.

Unlike the finite-difference or the finite-element volume-discretization methods, SIE methods treat infinite homogeneous regions (and some other cases) analytically via Green's functions, and therefore often require no artificial truncation of space. Because SIE methods only require surfaces to be discretized, they can be computationally efficient for problems involving piecewise homogeneous media. A popular SIE method is the boundary-element method (BEM) \cite{harrington89boundary, umashankar86electromagnetic, rao82electromagnetic, wang98a, sheng98solution}, particularly since the development of fast $O(N\mbox{log}N)$ solvers \cite{Phillips97a, Zhu03algorithms,zhang05rcs,song97multilevel}. However, a truncation difficulty arises with unbounded surfaces of infinitely extended channels common in photonics. Fig.~\ref{fig:device} is a general photonic device schematic with input and output waveguide channels. In order to accurately simulate and characterize the device, such as calculating its scattering parameters, formally, the transmission channel must be extended to infinity, requiring infinite computational resources. A more realistic option is to truncate the domain with an absorber that does not generate reflections.

The key challenge is to design an absorber that both has small reflections and is also easily incorporated into an SIE solver. The best-known absorber is a perfectly matched layer (PML) \cite{berenger94a,chew94a,sacks95a,gedney96an,Taflove95compuatational}, but a PML corresponds to a continuously varying anisotropic absorbing medium, whereas SIE methods are designed for piecewise homogeneous media. A similar problem arises if one were to simply add some absorption within the waveguides---in order to minimize transition reflection, the absorption would need to increase gradually from zero \cite{oskooi08the}, corresponding again to inhomogeneous media. One could also use a volume integral equation (VIE) \cite{schaubert84a} or a hybrid finite-element method in the inhomogeneous absorbing region, but then one would obtain numerical reflections from the discontinuous change in the discretization scheme from the SIE to the VIE. Moreover, it has been proposed that an integral-equation PML can be obtained by varying the Green's function instead of the media \cite{Pissoort03termination}, but a continuously varying Green's function greatly complicates panel integrations and fast solvers for SIE methods.

\begin{figure}[tb]
\begin{center}
\epsfclipon
\scalebox{0.5} {\epsffile{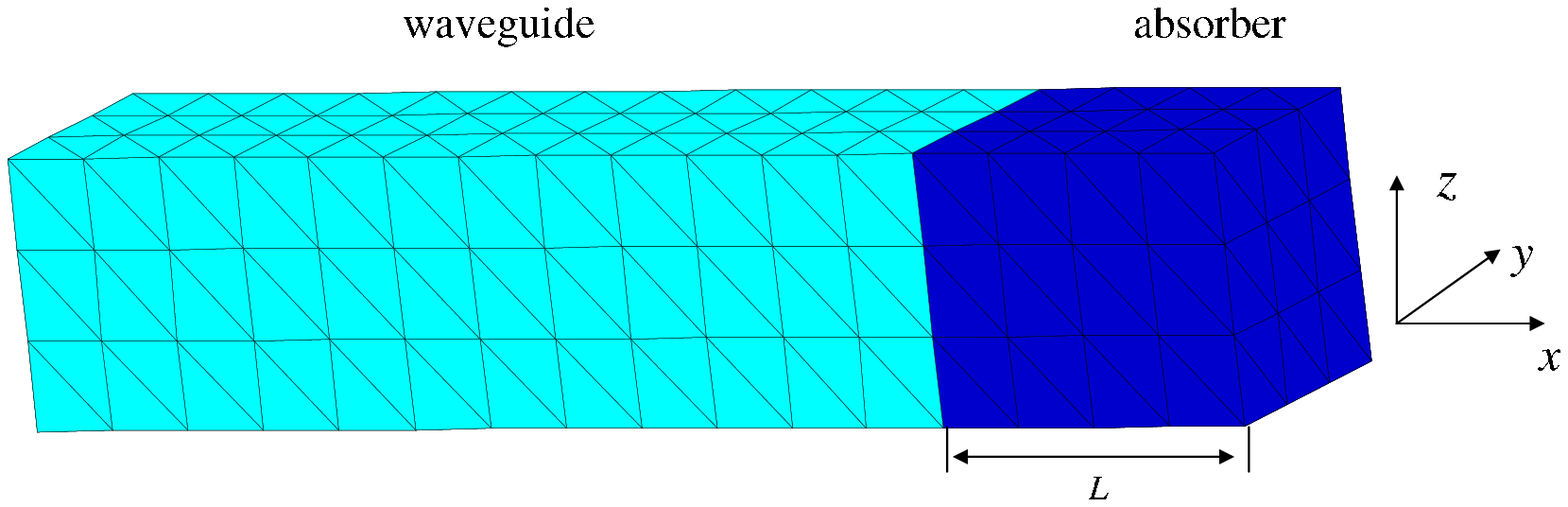}}
\caption{A discretized dielectric waveguide with an absorber attached.}
\label{fig:waveguide}
\end{center}
\end{figure}

\begin{figure}[!tb]
\begin{center}
\epsfclipon
 \subfigure[A waveguide with a \emph{volume} absorber]
 {\label{fig:wg_cross_volume}
 \scalebox{0.5}
 {\epsffile{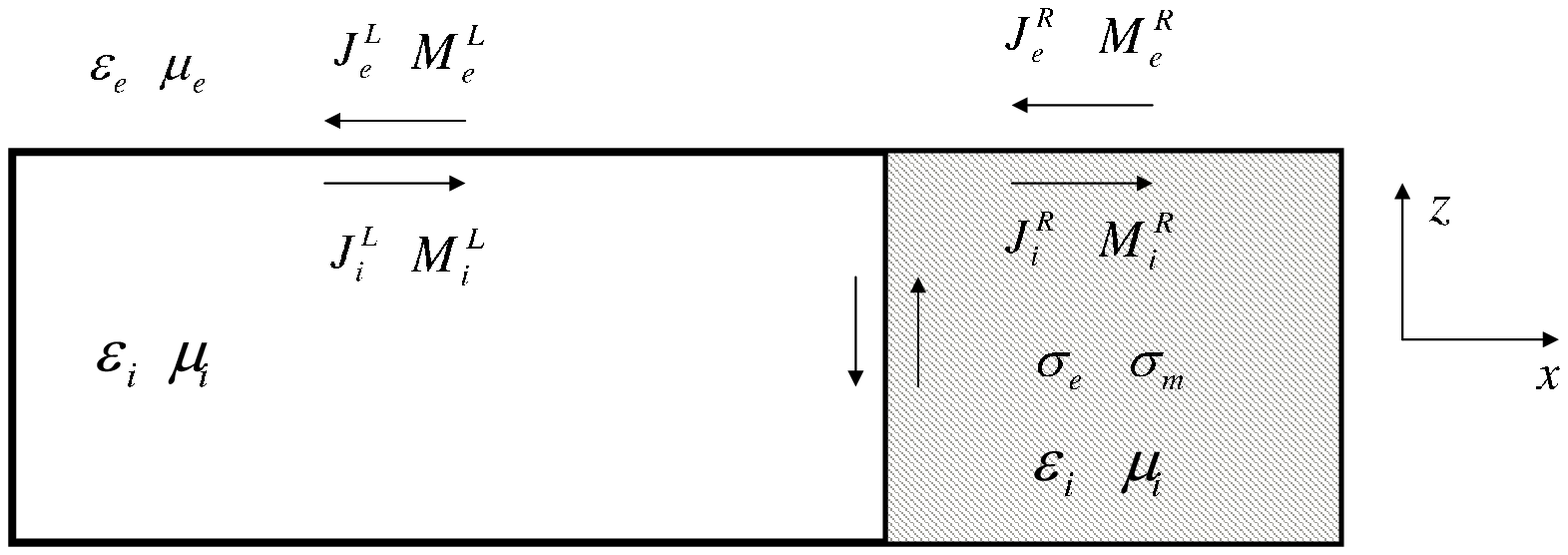}}
 }
\vspace{0.5cm}
  \subfigure[A waveguide with a \emph{surface} absorber]
 {\label{fig:wg_cross_surface}
 \scalebox{0.5}
 {\epsffile{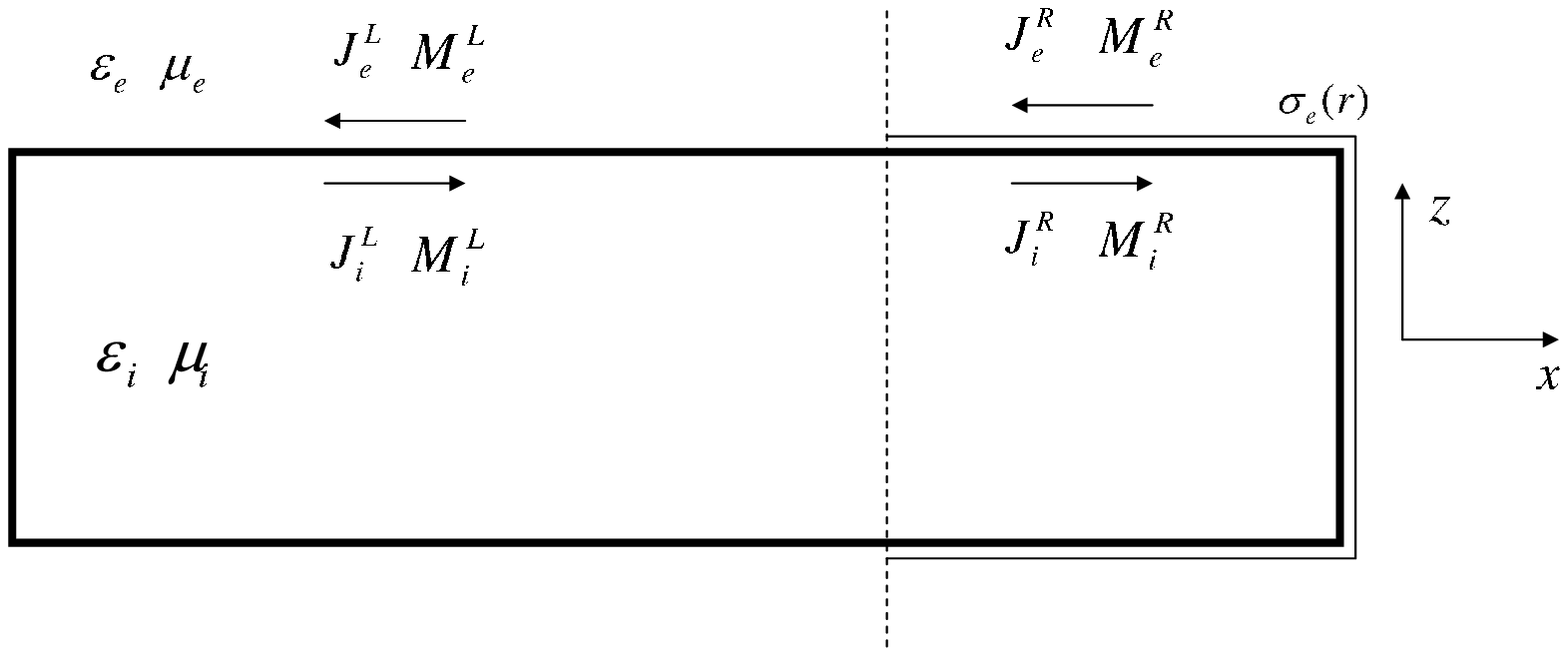}}
 }
\caption{The 2-D longitudinal section of a waveguide with an absorber. The lengths of the waveguide and absorber are $20\lambda_i$ and $10\lambda_i$, respectively, with $\lambda_i$ denoting the wavelength in the waveguide medium. The waveguide cross section size is $0.7211\lambda_i\times0.7211\lambda_i$. The relative permittivities of the waveguide (silicon) and the external medium (air) are $11.9$ and $1$, respectively.}
\label{fig:WG_long_cross}
\end{center}
\end{figure}

With SIE methods, one could, in principle, implement a scalar waveguide absorption in a piecewise homogeneous fashion as a discontinuous increase in absorption, but this will obviously generate large reflections due to the discontinuity of the medium and also the numerical reflection due to the discretization of the interfaces. We demonstrate this with a finite rectangular waveguide, to which an absorber with constant volume electrical conductivity $\sigma_e$ and magnetic conductivity $\sigma_m$ is attached, as shown in Fig.~\ref{fig:waveguide}. The longitudinal cross-section of this arrangement is shown in Fig.~\ref{fig:wg_cross_volume}. To achieve small reflections, the intrinsic impedance of the absorber is matched to that of the waveguide. Hence, the volume electrical conductivity $\sigma_e$ and the volume magnetic conductivity $\sigma_m$ should satisfy $\sigma_e/\sigma_m = \epsilon_i/\mu_i$, where $\epsilon_i$ and $\mu_i$ are the permittivity and permeability of both the waveguide and absorber media, respectively. We quantify the reflection by use of the standing wave ratio (SWR), the ratio of the maximum field magnitude to the minimum field magnitude in the standing-wave region, evaluated on the waveguide axis. From the SWR, a reflection coefficient is then readily obtained as in a conventional transmission line. We calculated the field reflection coefficients in this way for a range of $\sigma_e$ and $\sigma_m$, and the smallest reflection coefficient obtained was $12\%$. This value, which is listed in the first column of Table~\ref{table:reflection}, is unacceptable for many design applications. In particular, the design of tapers \cite{povinelli05slow} requires field reflection from terminations in the order of $10^{-3}$ or smaller. Evidently, a more sophisticated way of terminating waveguides is called for.

In this paper, we examine an alternative approach to absorbers, adding electrical conductivity to the waveguide \emph{surface} rather than to the volume, via a delta-function conductivity on the absorber surface, as shown in Fig.~\ref{fig:wg_cross_surface}. The absorber's interior medium remains the same as the waveguide's, thus eliminating the need to discretize the waveguide-absorber interface, shown as a solid line in Fig.~\ref{fig:wg_cross_volume} and a dashed line in Fig.~\ref{fig:wg_cross_surface}. This surface-conductivity strategy permits an efficient surface-only discretization, but at the same time allows for a smoothly increasing surface conductivity, thereby reducing transition reflections. Specifically, surface conductivity is easily implemented in SIE methods as it corresponds to a jump discontinuity in the field boundary conditions at the absorber surface. Since the SIE explicitly discretizes the surface boundary, continuously varying the field boundary conditions is easily implemented.

\begin{figure}[tbhp]
\begin{center}
  \subfigure[constant surface conductivity]
 {
 \scalebox{0.6}
 {\includegraphics{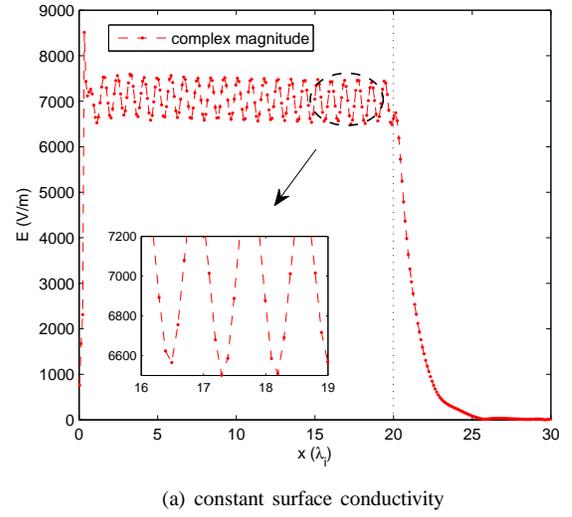}}
 }
 \subfigure[linear surface conductivity along $x$ direction]
 {
 \scalebox{0.6}
 {\includegraphics{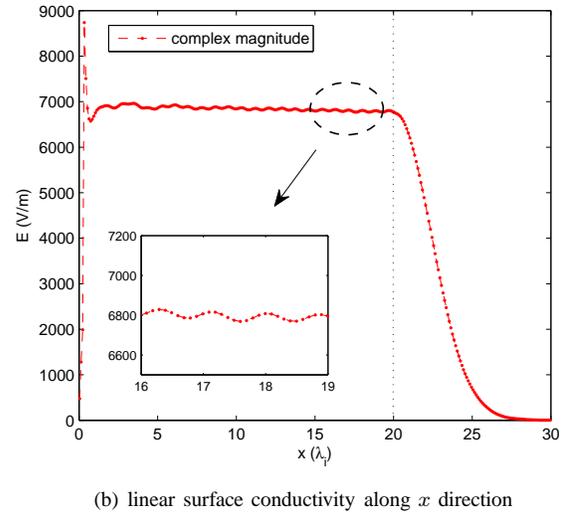}}
 }
 \subfigure[quadratic surface conductivity along $x$ direction]
 {
 \scalebox{0.6}{\includegraphics{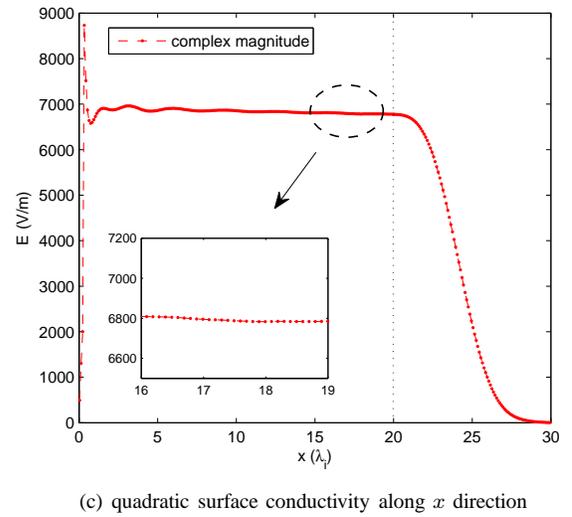}}
 }
 \caption{The complex magnitude of the electric field inside a waveguide and a surface absorber. The dashed line indicates the position of the waveguide-absorber interface.}
\label{fig:two_WG}
 \end{center}
\end{figure}

As a preview of results to be described below, in Fig.~\ref{fig:two_WG}, we show the numerically computed complex magnitudes of the electric field along the $x$ direction inside a rectangular waveguide with several different surface absorbers. The surface absorber is in the region where $x_0<x<(x_0+L)$ in which $x_0$ is the position of the interface and $L$ is the absorber length. The surface electrical conductivity in this region is given by $\sigma_e(x)=\sigma_0(\frac{x-x_0}{L})^d$, where $d=0,1,2$ for constant, linear and quadratic profiles. The constant $\sigma_0$ is chosen so that the total attenuation over the length of the absorbing region matches that of the optimal volume absorber above. The approach for calculating the attenuation along the absorber will be explained in section~\ref{reflection_class}.
In the complex magnitude plots of Fig.~\ref{fig:two_WG}, the peak-to-peak magnitudes of ripples are an indication of the amount of reflections. As is easily seen in Fig.~\ref{fig:two_WG}, there are substantial reflections when using a constant conductivity, smaller reflections when using a linearly increasing conductivity, and almost no reflections for a quadratically increasing conductivity. The magnitudes of the field reflection coefficients $r$ are listed in Table~\ref{table:reflection}, and show that the reflection coefficient for the quadratically varying surface conductivity is nearly one thousand times smaller than the reflection coefficient for the volume absorber. The results for the constant, linearly and quadratically varying surface absorber verify the results in \cite{oskooi08the}, that the smoothness of any transition in the waveguide largely determines the resulting reflection.

This paper is organized as follows. In section~\ref{sec:formulation}, the BEM formulation incorporating surface conductivity is derived. In section~\ref{sec:decay_rate}, the decay rate due to the surface conductivity is examined using both numerical experiments and calculations using perturbation theory and Poynting's theorem. In section~\ref{sec:absorber}, the asymptotic behavior of the transition reflection with respect to absorber length is presented. In section~\ref{sec:field}, we note the existence of radiation modes originating from the excitation source mismatch with the waveguide mode, and show that the radiation complicates the interpretation of certain numerical results, but does not affect the performance of the surface absorber. Details of the numerical implementation of the SIE solver are given in Appendix, specifically, the detailed matrix construction, acceleration and preconditioning techniques.

\begin{table}[tb]
\caption{The Standing wave ratio (SWR) and field reflection versus the conductivity
distribution of the absorber}
\begin{center}
\begin{tabular}{|c|c|c|c|c|}
\hline
Absorber type & Volume&\multicolumn{3}{|c|}{Surface} \\
\hline
Conductivity profile & Constant &Constant & Linear & Quadratic \\
\hline
SWR & $1.2933$ &$1.1477$ & $1.0062$ & $1.0003$ \\
\hline
Reflection $r$ & $0.1279$& $0.0688$ & $0.0031$ & $1.4998\times 10^{-4}$\\
\hline
\end{tabular}
\end{center}
\label{table:reflection}
\end{table}

\section{BEM formulations with the Surface Conductive Absorber}\label{sec:formulation}

In this section, we describe the 3-D BEM formulation for a waveguide truncated with a surface conductive absorber. Fig.~\ref{fig:wg_cross_surface} shows the $x$\textendash$z$ plane cross-section of an $x$-directed truncated rectangular waveguide, where the surface conductive absorber region is to the right of the dashed line. The permittivity and permeability of the waveguide interior and the exterior media are denoted as $\epsilon_e$, $\mu_e$ and $\epsilon_i$, $\mu_i$, where the subscripts $e$ and $i$ denote the exterior and the interior, respectively. The electrical surface conductivity $\sigma_e(\mathbf{r})$ is subscripted with $e$ as a reminder that only electrical conductivity is being considered, though the generalization of what follows to both electrical and magnetic conductivity could be considered. As is described in section \ref{sec:decay_rate}, using only electrical conductivity can have a saturation phenomenon that can be avoided at the cost of using a longer absorber. The system is excited by a Gaussian beam propagating in $+\hat{x}$ direction. The Gaussian beam is generated by a dipole in a complex space \cite{erez94electromagnetic}, where the real part of the dipole position is inside the waveguide, $\frac{1}{4}\lambda_i$ from the left end. In this paper, the convention of the $e^{j\omega t}$ time-harmonic mode is adopted.

In SIE methods, for computing time-harmonic solutions, the unknowns are surface variables. In our case, we use surface electric and magnetic currents on both the interior, $\mathbf{J}_i$ and $\mathbf{M}_i$, and exterior, $\mathbf{J}_e$ and $\mathbf{M}_e$, of every surface. The currents on surfaces with $\sigma_e=0$ on the left side of Fig.~\ref{fig:wg_cross_surface}, satisfy a simpler set of equations than the currents on surfaces where $\sigma_e\neq 0$, the right side of Fig.~\ref{fig:wg_cross_surface}. When appropriate, we distinguish between the $\sigma_e=0$ and $\sigma_e\neq 0$ currents with superscripts $L$ and $R$, respectively.

Invoking the equivalence principle \cite{harrington61time} yields a relation between surface currents and fields
\begin{eqnarray}
-\hat{n}\times \mathbf{E}_{e} &=& \mathbf{M}_e, \label{eq6} \\
\hat{n}\times \mathbf{H}_{e} &=& \mathbf{J}_e, \label{eq7}\\
\hat{n}\times(\mathbf{E}_{i}+\mathbf{E}_{\mathrm{inc}}) &=& \mathbf{M}_i, \label{eq8}\\
-\hat{n}\times(\mathbf{H}_{i}+\mathbf{H}_{\mathrm{inc}}) &=& \mathbf{J}_i, \label{eq9}
\end{eqnarray}
where $\hat{n}$ is an exterior-directed normal unit-vector,
$\mathbf{E}_{\mathrm{inc}}$, $\mathbf{H}_{\mathrm{inc}}$, are the electric and magnetic fields of the Gaussian beam in a homogeneous space with material parameters equal to those of the waveguide interior, and $\mathbf{E}_e$, $\mathbf{H}_e$ and $\mathbf{E}_i$, $\mathbf{H}_i$ are electric and magnetic fields due to the equivalent currents in the exterior and interior, respectively.

On the surface of the waveguide, the continuity of the tangential components of the electric and magnetic fields yields the well-known PMCHW formulation \cite{harrington89boundary, umashankar86electromagnetic},
\setlength\arraycolsep{0pt}\begin{eqnarray}
\hat{n}\times \mathbf{E}^{L}_{e}&(& \mathbf{J}^L_e, \mathbf{M}^L_e,
\mathbf{J}^R_{e}, \mathbf{M}^R_e ) \nonumber \\
&=& \hat{n}\times[\mathbf{E}^{L}_{i}( \mathbf{J}^L_i, \mathbf{M}^L_i,
\mathbf{J}^R_{i}, \mathbf{M}^R_i ) +\mathbf{E}^{L}_{\mathrm{inc}}], \label{eq1} \\
\hat{n}\times \mathbf{H}^{L}_{e}&(& \mathbf{J}^L_e, \mathbf{M}^L_e,
\mathbf{J}^R_{e}, \mathbf{M}^R_e ) \nonumber
\\&=& \hat{n}\times[\mathbf{H}^{L}_{i}( \mathbf{J}^L_i, \mathbf{M}^L_i,
\mathbf{J}^R_{i}, \mathbf{M}^R_i )+\mathbf{H}^{L}_{\mathrm{inc}}], \label{eq2}
\end{eqnarray}
where $\mathbf{E}^L(\cdot)$ and $\mathbf{H}^L(\cdot)$ are integral operators described in the Appendix. From (\ref{eq6})-(\ref{eq9}) and the tangential field continuity in the surface conductivity free region, the equivalent currents on the two sides of the waveguide surface are of the equal magnitude, but are opposite in direction. Specifically,
\setlength\arraycolsep{1pt}
\begin{eqnarray}
  \mathbf{J}_e^L~ =& -\mathbf{J}_i^L& = ~\mathbf{J}^L, \label{eq10} \\
  \mathbf{M}_e^L~ =& -\mathbf{M}_i^L& = ~\mathbf{M}^L. \label{eq11}
\end{eqnarray}
Thus, the unknown currents on the waveguide side are reduced to $\mathbf{J}^L$ and $\mathbf{M}^L$.

For the surfaces where $\sigma_e\neq 0$, a modified surface formulation is needed, one that incorporates the discontinuity due to the surface conductivity. When $\sigma_e\neq 0$, the
tangential electric field is still continuous across the absorber surface, and therefore
\setlength\arraycolsep{0pt}
\begin{eqnarray}
 \hat{n}\times \mathbf{E}^{R}_{e}&(& \mathbf{J}^L_e, \mathbf{M}^L_e,
\mathbf{J}^R_{e}, \mathbf{M}^R_e ) \nonumber\\
&=&\hat{n}\times[\mathbf{E}^{R}_{i}( \mathbf{J}^L_i, \mathbf{M}^L_i,
\mathbf{J}^R_{i}, \mathbf{M}^R_i )+\mathbf{E}^{R}_{\mathrm{inc}}]. \label{eq3}
\end{eqnarray}
The tangential magnetic field is not continuous as a sheet of surface electric current $\mathbf{J}_{\mathrm{ind}} =\sigma_e\mathbf{E}_{\mathrm{tan}}^R$ is induced due to the electrical surface conductivity, thus creating a jump. Therefore,
 \setlength\arraycolsep{0pt}
\begin{eqnarray}
 \hat{n}\times &[& \mathbf{H}^{R}_{e}(\mathbf{J}^L_e, \mathbf{M}^L_e,
\mathbf{J}^R_{e}, \mathbf{M}^R_e ) \hspace{2cm}\nonumber\\
&&- \mathbf{H}^{R}_{i}( \mathbf{J}^L_i, \mathbf{M}^L_i,
\mathbf{J}^R_{i}, \mathbf{M}^R_i )-\mathbf{H}^{R}_{\mathrm{inc}}] = \sigma_e\mathbf{E}_{\mathrm{tan}}^{R},
\label{eq4}
\end{eqnarray}
where $\mathbf{E}_{\mathrm{tan}}^{R}=-\hat{n}\times(\hat{n}\times\mathbf{E}^R)$ is the tangential electric field on the absorber surface, and could choose the field on either side of the absorber surface according to the enforced equality in (\ref{eq3}).

As a result of (\ref{eq6}), (\ref{eq8}) and (\ref{eq3}), the interior and exterior magnetic currents can be represented with a single variable
\begin{equation}
  \mathbf{M}_e^R = -\mathbf{M}_i^R = \mathbf{M}^R, \label{eq12}
\end{equation}
but the discontinuity of tangential magnetic field implies $|\mathbf{J}^{R}_{e}| \neq |\mathbf{J}^{R}_{i}|$, and (\ref{eq7}), (\ref{eq9}), and (\ref{eq4}) must be combined to generate a local equation
\begin{equation}
  \mathbf{J}^R_{e} + \mathbf{J}^R_{i} = \sigma_e\mathbf{E}_{\mathrm{tan}}^{R}.\label{eq5}
\end{equation}
Finally, using the integral operator relation between $\mathbf{E}_{\mathrm{tan}}^{R}$ and $\mathbf{J}$ and $\mathbf{M}$, and substituting into (\ref{eq4}), (\ref{eq5}), yields
{\small\setlength\arraycolsep{0pt}
\begin{eqnarray}
 \hat{n}\times [\mathbf{H}^{R}_{e}( \mathbf{J}^L_e&,&
 \mathbf{M}^L_e,\mathbf{J}^R_{e}, \mathbf{M}^R_e ) - \mathbf{H}^{R}_{i}(
 \mathbf{J}^L_i, \mathbf{M}^L_i, \mathbf{J}^R_{i}, \mathbf{M}^R_i )\nonumber \\
&&-\mathbf{H}^{R}_{\mathrm{inc}}]
= \sigma_e\mathbf{E}^{R}_{e,\mathrm{tan}}( \mathbf{J}^L_e, \mathbf{M}^L_e,
\mathbf{J}^R_{e}, \mathbf{M}^R_e ), \label{eq13}\\
\mathbf{J}^R_{e} + \mathbf{J}^R_{i} =
\sigma_e[&& \mathbf{E}^{R}_{i,\mathrm{tan}}(
\mathbf{J}^L_i, \mathbf{M}^L_i, \mathbf{J}^R_{i}, \mathbf{M}^R_i
)+\mathbf{E}^{R}_{\mathrm{inc},\mathrm{tan}}].\label{eq14}
\end{eqnarray}}
\setlength\arraycolsep{5pt}

The unknowns, the surface currents $\mathbf{J}^L$, $\mathbf{M}^L$, $\mathbf{J}_{e}^R$, $\mathbf{J}_{i}^R$ and $\mathbf{M}^R$, can be determined by solving equations (\ref{eq1}), (\ref{eq2}) on the left waveguide surfaces and (\ref{eq3}), (\ref{eq13}), (\ref{eq14}) on the right absorber surfaces. The resulting linear system can be constructed and solved as described in Appendix. It is possible to simplify the formulation by eliminating one extra variable of electric currents on the absorber surface, though there are some accuracy issues \cite{zhang10calculation}.

\section{The decay rate of the fields} \label{sec:decay_rate}

\begin{figure}[tb]
\begin{center}
\scalebox{0.5} {\includegraphics{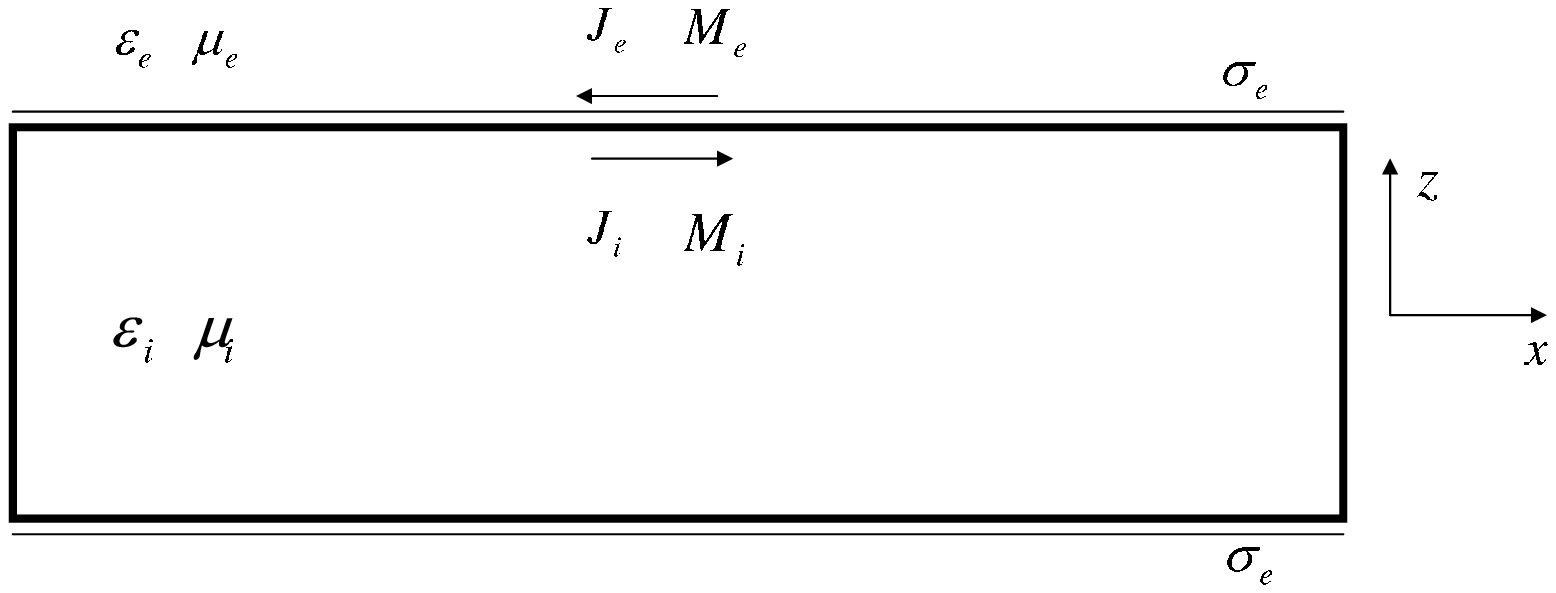}}
\caption{The 2-D longitudinal section of a waveguide with uniform surface conductivity. The waveguide length is $10\lambda_i$ and cross section size is $0.7211\lambda_i\times0.7211\lambda_i$. The relative permittivity of the waveguide and external medium are 11.9 and 1, respectively.}
\label{fig:one_wg_cross}
\end{center}
\end{figure}

In this section the exponential decay rate of waves propagating through the surface absorber region is analyzed. We demonstrate the relation between decay rate and surface conductivity using the example of a single dielectric waveguide with uniformly distributed surface conductivity. The longitudinal cross-section is shown in Fig.~\ref{fig:one_wg_cross}. The behavior of interior fields generated by a Gaussian beam source were computed using a BEM method based on solving (\ref{eq3}), (\ref{eq13}) and (\ref{eq14}).

\begin{figure}[tb]
\begin{center}
\scalebox{0.6} {\includegraphics{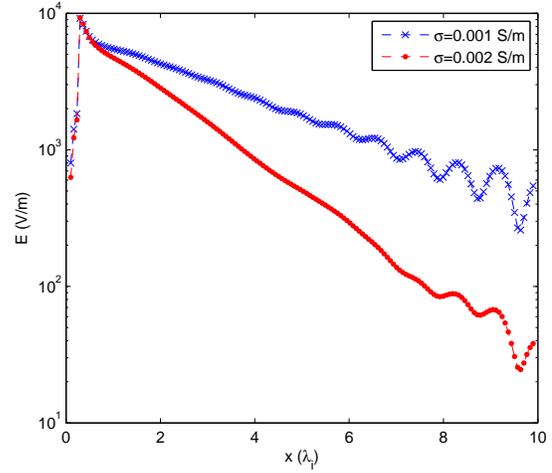}}
\caption{The complex magnitude of the electric field along $x$ inside the
waveguide in Fig.~\ref{fig:one_wg_cross} with uniform surface conductivity.} \label{fig:one_WG}
\end{center}
\end{figure}

The plots in Fig.~\ref{fig:one_WG} show the complex magnitudes of the electric fields along the $x$ axis inside the waveguide for two cases, $\sigma_e=0.001$ S and $\sigma_e=0.002$ S. As expected, the complex magnitude decays exponentially with distance from the source with a surface-conductivity dependent rate. Also, as can be seen, waves reflect back from the right end and presumably these reflections decay as they travel to the left.

\begin{figure}[!tb]
\begin{center}
\scalebox{0.6} {\includegraphics{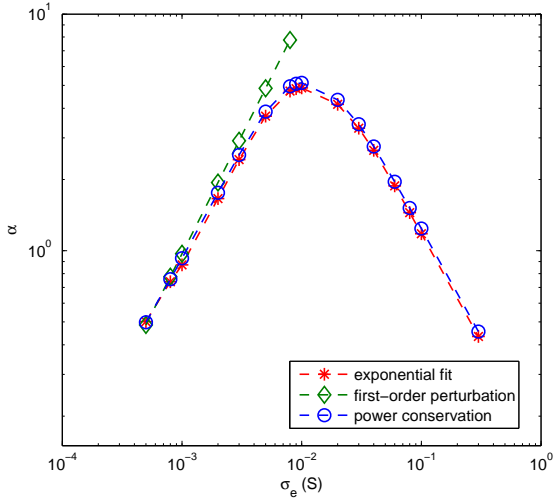}}
\caption{A comparison of three methods for computing the rate of field exponential decay along the propagation direction versus surface conductivity.} \label{fig:decay_rate}
\end{center}
\end{figure}

An approximation to the rate of exponential decay can be determined by fitting the field plots. The fitted decay rates for a range of $\sigma_e$ are shown in Fig.~\ref{fig:decay_rate} and denoted with a dashed star curve. The decay rate does not monotonically increase with the surface conductivity. The curve shape can be explained as follows. When $\sigma_e$ is small, the propagating wave is able to penetrate the lossy surface and is absorbed, with the absorption increasing with $\sigma_e$ as expected. However, for large $\sigma_e$, the surface conductor itself becomes reflecting, forming essentially an enclosed metallic waveguide; as $\sigma_e\rightarrow\infty$ the tangential electric field vanishes at the surface and therefore there is no absorption. The practical implications of this upper bound on effective values for $\sigma_e$ are limited, and are described in section~\ref{sec:absorber}.

The following sections introduce two alternative approaches to calculate the decay rate from surface conductivities, to confirm and further illustrate the numerical observations above.

\subsection{Decay rate calculation by perturbation theory}
In this section, a first-order closed-form decay rate formula, valid for low surface conductivity, is derived using perturbation theory.

Assume the electric field $\mathbf{E}^{(0)}(\mathbf{r})$ of the fundamental mode of a lossless rectangular waveguide is given, and the superscript $(0)$ denotes the unperturbed quantity. When electrical surface conductivity $\sigma_e$ is put on the waveguide surface, it is equivalent to a perturbation of permittivity, denoted as $\Delta\epsilon(\mathbf{r})=-\frac{j\sigma_e}{\omega}\delta_S(\mathbf{r})$, where $\delta_S$ is the Dirac delta function across the waveguide surface. According to \cite{harrington61time}, \cite{joannopoulos08photonic}, the first-order variance of angular frequency due to the perturbation of permittivity $\Delta\epsilon(\mathbf{r})$ is
\begin{equation}\label{eq40}
 \Delta\omega^{(1)}=-\frac{\omega}{2}\frac{\int_V \Delta\epsilon(\mathbf{r})|\mathbf{E}^{(0)}(\mathbf{r})|^2dV}{\int_V\epsilon|\mathbf{E}^{(0)}(\mathbf{r})|^2dV},
\end{equation}
where $V$ is the whole volume domain and the superscript $(1)$ denotes a first-order approximation. After applying the triple product rule to partial derivatives of the three interdependent variables $\omega$, $k$, $\sigma_e$, we obtain a first-order change in propagation constant $k$ due to the frequency change in (\ref{eq40}), denoted as $\Delta k^{(1)} = -\frac{\Delta\omega}{V_g}$, where $V_{g}$ is the group velocity, as $V_g=\frac{\partial\omega}{\partial k}\rvert_{\sigma_e}$ in which the subscript indicates $\sigma_e$ is held fixed. Combining (\ref{eq40}) and equations of $\Delta\epsilon(\mathbf{r})$ and $\Delta k^{(1)}$ above, the integral in the numerator of (\ref{eq40}) is reduced to a surface integral of the tangential components of the electric field, therefore
\begin{equation}\label{eq41}
 \Delta k^{(1)} =-\frac{j\sigma_e}{2V_g}\frac{\int_S|\mathbf{E}_{\mathrm{tan}}^{(0)}(\mathbf{r})|^2 dS}{\int_V\epsilon|\mathbf{E}^{(0)}(\mathbf{r})|^2dV},
\end{equation}
where $S$ is the surface of the waveguide. As expected, the perturbation in the propagation constant is imaginary, which in turn represents the decay rate $\alpha^{(1)}=-\mbox{Im}\{\Delta k^{(1)}\}$. With a uniform cross section, the volume and surface integrals in (\ref{eq41}) can be further reduced to surface and line integrals on the cross section, respectively. The electric field before the perturbation $\mathbf{E}^{(0)}(\mathbf{r})$, along with $V_g$, can be obtained numerically, for example, with a planewave method \cite{johnson01block}.

The decay rate calculated using the perturbation is plotted in Fig.~\ref{fig:decay_rate} with a dashed diamond curve. Note that the curve overlaps with decay rates computed with other methods when the surface conductivity is small, and deviates for larger conductivity as should be expected given the first-order approximation.

\subsection{Decay rate calculation using Poynting's theorem}

\begin{figure}[tb]
\begin{center}
\scalebox{0.55} {\includegraphics{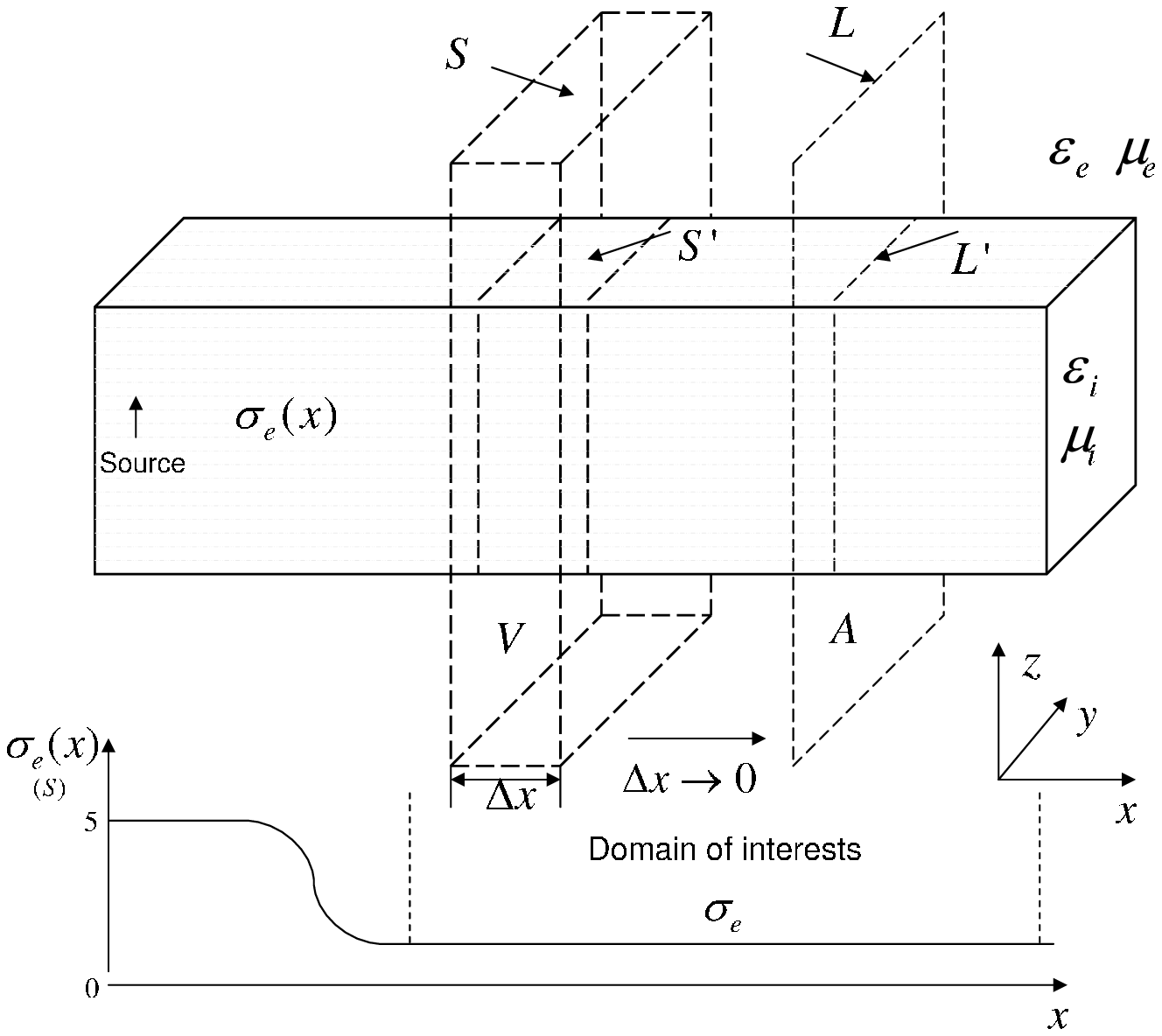}}
\caption{Illustration of the approach using Poynting's theorem to calculate the decay rate of a waveguide with surface conductivity. The plot of the surface conductivity distribution $\sigma_e(x)$ along the longitudinal direction is aligned with the waveguide.}
\label{fig:one_wg_cross_poynting}
\end{center}
\end{figure}

The perturbation method above predicts the decay rate when the surface conductivity is small. An alternative approach, based on Poynting's theorem, can be used to calculate the decay rate for the entire $\sigma_e$ range.

Figure~\ref{fig:one_wg_cross_poynting} shows a waveguide with surface conductivity and also plots the conductivity function $\sigma_e(x)$ with $x$-axis aligned with the $x$-axis of the waveguide. Since this approach requires integrating the fields of source-excited propagating modes in the exterior region, some inevitably excited modes, such as radiation modes that will be discussed in section \ref{sec:field}, must be suppressed. For this reason, the surface conductivity starts with a large constant value ($5$ S). The large surface conductivity leads to the saturation as seen in Fig.~\ref{fig:decay_rate}, and therefore, the Gaussian-beam source excites metallic waveguide modes at the beginning, propagating in $+\hat{x}$ direction in the closed interior region. The surface conductivity is then smoothly reduced to a smaller value, with which the decay rate is to be calculated. In this way, the metallic waveguide modes can smoothly change to the desired decaying dielectric waveguide modes with minimal radiation modes excited.

The decaying propagating mode in the domain of interest, shown in Fig.~\ref{fig:one_wg_cross_poynting}, is assumed in the form
\begin{eqnarray}
 \mathbf{E}(\mathbf{r})&\sim& \mathbf{E_0}(y,z)e^{-\alpha x -j\beta x}, \label{eq_Eprop}\\
 \mathbf{H}(\mathbf{r})&\sim& \mathbf{H_0}(y,z)e^{-\alpha x -j\beta x},  \label{eq_Hprop}
\end{eqnarray}
where $\beta$ is the real propagation constant, and $\alpha$ is the unknown decay rate. The Poynting vector in frequency domain is given by $\mathbf{S}=\frac{1}{2}\mathbf{E}\times \mathbf{H}^*$, and together with the assumed forms of $\mathbf{E}$ in (\ref{eq_Eprop}) and $\mathbf{H}$ in (\ref{eq_Hprop}), the derivative with respect to $x$ is given by
\begin{equation}\label{eq_S_derivative}
\frac{d\mathbf{S}}{dx}=-2\alpha \mathbf{S}.
\end{equation}
As illustrated in Fig.~\ref{fig:one_wg_cross_poynting}, apply Poynting's theorem in the closed volume $V$
\begin{equation}\label{eq_poynting_thm}
\mbox{Re} \int_S\mathbf{S}\cdot\hat{n}dS = -\frac{1}{2}\int_{S'}\sigma_e|\mathbf{E_{\mathrm{tan}}}|^2 dS',
\end{equation}
where $S$ is the surface of the volume $V$, $\hat{n}$ is an exterior-directed normal unit-vector, and $S'$ is the waveguide surface within $V$. In the limit as $\Delta x \rightarrow 0$, the closed integration surface $S$ becomes a  surface $A$, and one component of the integrand of the left side of (\ref{eq_poynting_thm}) becomes $\frac{d\mathbf{S}}{dx}\cdot\hat{x}$. Combining (\ref{eq_S_derivative}) and this $\Delta x \rightarrow 0$ limit of (\ref{eq_poynting_thm}) yields a closed-form representation of the decay rate $\alpha$
\begin{equation}\label{eq_alpha_power}
\alpha=\frac{\int_{L'}\sigma_e |\mathbf{E_{\mathrm{tan}}}|^2 dl' + \mathrm{Re}\int_L (\mathbf{E}\times \mathbf{H}^*)\cdot\hat{n}dl}{2Re\int_A(\mathbf{E}\times \mathbf{H}^*)\cdot\hat{x}dS},
\end{equation}
where $L$ denotes the boundary of the surface $A$, and $L'$ denotes the integral line on the waveguide surface within the surface $A$.

The decay rate calculated using (\ref{eq_alpha_power}) is plotted in Fig.~\ref{fig:decay_rate} with a dashed circle curve. It shows good agreement with the decay rate computed using fitting for the entire $\sigma_e$ range, verifying that the surface conductivity is handled correctly by the BEM in accordance with Maxwell's equations.

\section{Reflections and Asymptotic Convergence} \label{sec:absorber}

In section~\ref{sec:introduction}, we showed that a smoothly varying surface conductive absorber easily implemented in the SIE method is orders of magnitude more effective at eliminating reflections than a volume absorber of comparable length. And, since computational cost increases with the length of the absorber, it is worth examining the relationship between reflections and the absorber length. In order to do this, we first require a more careful classification of reflections and a more sensitive numerical measure of reflections than the standing wave ratio method. In the following subsections, we will define round-trip and transition reflections and present a measure of the power-law asymptotic convergence of transition reflections.

\subsection{Round-trip and transition reflections}\label{reflection_class}
Reflections in a domain of interest can be divided into a round-trip reflection, $R_r$, and a transition reflection, $R_t$. The round-trip reflection is generated by waves entering into the absorber, propagating to the end without being completely absorbed, reflected off the end of the absorber, and eventually propagating back into the waveguide. The round-trip reflection coefficient is proportional to
\begin{equation}\label{eq_roundtrip}
 R_r = Ce^{-4\int_{L}\alpha(x)dx},
\end{equation}
where the coefficient $C$ is determined by the reflection at the end of the absorber and $L$ is the absorber length. A factor of $2$ in the exponent of (\ref{eq_roundtrip}) represents the effect of the round trip, and another factor of $2$ indicates that the power is considered.

The transition reflection is the reflection generated by the change in material properties at the waveguide-absorber interface. Therefore, smoother material change at the interface produces smaller transition reflection as predicted by coupled-mode theory \cite{oskooi08the}.

If the round-trip reflection $R_r$ in (\ref{eq_roundtrip}) is held fixed, the decay rate $\alpha$ is inversely related to the absorber length $L$. In another words, for a longer absorber, a smaller decay rate $\alpha$ yields the same round-trip reflection. Since $\alpha$ is proportional to the surface conductivity $\sigma_e$ for small $\sigma_e$, using a longer absorber implies a smaller transition reflection. It was further shown in \cite{oskooi08the} that, given a fixed round-trip reflection, the transition reflection decreased as a power law with increasing absorber length $L$. The power-law exponent is determined by the order of the differentiability of the medium (conductivity) function. Suppose a conductivity function along $x$ is $\sigma_e(x)=\sigma_{0}u(\frac{x-x_0}{L})$, where the interface is at $x=x_0$, and $u(s)$ is a monomial function with order $d$ in the $0\leq s \leq 1$ region
\setlength\arraycolsep{5pt}
\begin{equation} \label{eq60}
u(s)=\left\lbrace
\begin{array}{cc}
 s^d & 0\leq s \leq 1\\
0 & s<0
\end{array}
 \right..
\end{equation}
With the surface conductivity function $\sigma_e(x)$ and a fixed round-trip reflection, the asymptotic behavior of the transition reflection in terms of the length of the absorber is
\begin{equation}\label{eq43}
 R_t(L) \sim O\left(\frac{1}{L^{2d+2}}\right).
\end{equation}
The power-law behavior in (\ref{eq43}) indicates that, with a higher-order conductivity function, the transition reflection decreases faster with increasing the absorber length. It does not follow that $d$ should be made arbitrarily large, however, there is a tradeoff in which increasing $d$ eventually delays the onset $L$ of the asymptotic regime in which (24) is valid \cite{oskooi08the}.

\subsection{Asymptotic Convergence with $L$}
We present numerical results to verify the asymptotic power-law convergence of the transition reflection of the surface absorber. Because it is hard to explicitly measure the transition reflection in the integral equation method, instead, we measure alternative electric field expressions as in \cite{oskooi08the}.

\begin{figure}[tbhp]
\begin{center}
  \subfigure[The asymptotic convergence of the transition reflection measured using $\frac{|\mathbf{E}(L+1)-\mathbf{E}(L)|^2}{|\mathbf{E}(L)|^2}$.]
 {
 \scalebox{0.6}
 {\includegraphics{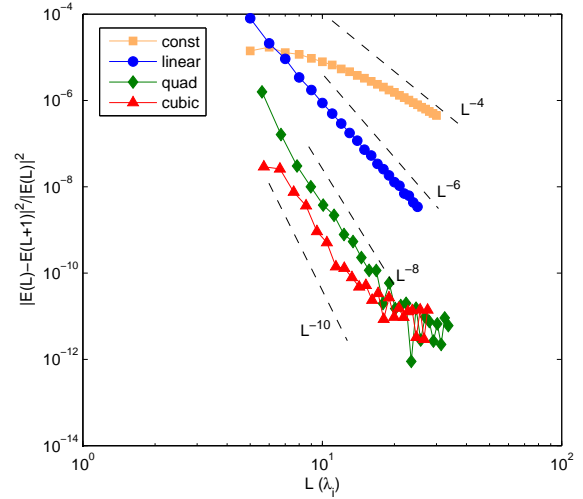}}\label{fig:convergence_unit_length}
 }
 \subfigure[The asymptotic convergence of the transition reflection measured using $\frac{|\mathbf{E}(2L)-\mathbf{E}(L)|^2}{|\mathbf{E}(L)|^2}$.]
 {
 \scalebox{0.6}
 {\includegraphics{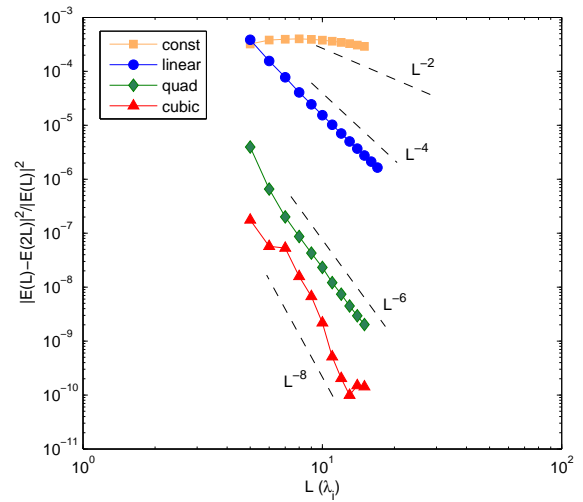}}\label{fig:convergence_double_length}
 }
 \caption{Asymptotic power-law convergence of the transition reflection with the length of the surface absorber. The length of the waveguide is $10\lambda_i$, with $\lambda_i$ denoting the wavelength in the waveguide medium. The waveguide cross section size is $0.7211\lambda_i\times0.7211\lambda_i$. The relative permittivities of the waveguide (silicon) and the external medium (air) are $11.9$ and $1$, respectively.}
 \end{center}
\end{figure}

First, we define $\mathbf{E}(L)$ as the electric field at a fixed position in the waveguide when the length of the absorber is $L$, with unit $\lambda_i$. Thus $\mathbf{E}(L)$ includes the incident field, the round-trip reflection and the transition reflection. With a small fixed round-trip reflection, the difference of $\mathbf{E}(L+1)$ and $\mathbf{E}(L)$ is the difference of the transition reflections, which in the limit of large $L$ approaches to zero, so $\mathbf{E}(L+1)-\mathbf{E}(L)\rightarrow 0$. Therefore, $|\mathbf{E}(L+1)-\mathbf{E}(L)|$, and similarly $|\mathbf{E}(2L)-\mathbf{E}(L)|$, can be a measure of transition reflection, and specifically, derived from (\ref{eq43}), they are subject to the following asymptotic behavior
\begin{eqnarray}
 \frac{|\mathbf{E}(L+1)-\mathbf{E}(L)|^2}{|\mathbf{E}(L)|^2} &\sim& O\left(\frac{1}{L^{2d+4}}\right),\label{eq44}\\
\frac{|\mathbf{E}(2L)-\mathbf{E}(L)|^2}{|\mathbf{E}(L)|^2} &\sim& O\left(\frac{1}{L^{2d+2}}\right).\label{eq45}
\end{eqnarray}
In the example of a rectangular waveguide attached with a surface absorber, the asymptotic convergence of (\ref{eq44}) is shown in Fig.~\ref{fig:convergence_unit_length} in a log-log scale for constant, linear, quadratic, and cubic surface conductivity profiles ($d=0,1,2,3$). The curve for $d=0$ slowly approaches to the $L^{-4}$ curve. The other three curves align with the expected $L^{-6}, L^{-8}, L^{-10}$ curves respectively. The curves would eventually converge to a small quantity, which is the difference of the small round-trip reflections due to a phase difference. Fig.~\ref{fig:convergence_double_length} shows the asymptotic convergence of (\ref{eq45}) with the same waveguide example for constant, linear, quadratic, and cubic profile surface conductivities. Again, the curve for the constant conductivity profile converges slowly, and is expected to be aligned with the $L^{-2}$ curve. The other three power-law convergence are the same as the predicted $L^{-4}, L^{-6}, L^{-8}$. The agreement of the figures verifies the power-law behavior of the transition reflection of the surface absorber and further illustrates that a higher-order conductivity function leads to smaller transition reflections.

\section{Radiation in the surface absorber}\label{sec:field}

\begin{figure}[tb]
\begin{center}
\scalebox{0.6} {\includegraphics{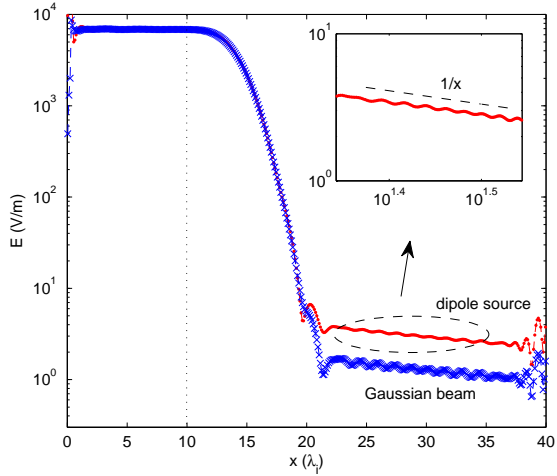}}
\caption{The complex magnitude of the electric field inside a waveguide and a long surface absorber excited by a dipole source and a Gaussian beam, respectively. The lengths of the waveguide and the absorber are $10\lambda_i$ and $30\lambda_i$, respectively. The surface conductivity on the absorber increases quadratically.} \label{fig:radiation_mode}
\end{center}
\end{figure}

A surface absorber with a conductivity that rises smoothly with distance from the waveguide-absorber interface would be expected to have interior fields whose magnitude decays with distance at an accelerating exponential rate. Instead, the field magnitudes decay exponentially near the interface, but then switch to an $O(\frac{1}{r})$ decay, as shown in Fig.~\ref{fig:radiation_mode}. The cause of this switch in decay rate is due to coupled radiation. The radiation is generated by the inevitable mismatch between the excitation source and waveguide modes. The situation is similar to a source radiating in a lossy half-space, in which the dominant field contribution is due to a lateral wave that decays algebraically \cite{wcchew:1990, ishimaru91electromagnetic}. Fig.~\ref{fig:radiation_mode} shows the complex magnitude of the interior electric field in a waveguide attached with a long quadratic-profile surface absorber. The waveguide system is excited by a dipole source and a Gaussian beam, respectively, located inside the waveguide. In the semilog plot, the dominant waveguide mode decays at an accelerating exponential rate, and then because the guided mode decays faster than the $O(\frac{1}{r})$ radiation, the radiation dominates at a certain distance from the waveguide-absorber interface. The coupled radiation results in an $O(\frac{1}{r})$ floor, as is clearly shown in the inset with a log-log scale. Note that using a Gaussian beam source results in a lower floor than using a dipole source, because the Gaussian beam is more directional, and generates less radiation.

The coupled $O(\frac{1}{r})$ radiation does not affect the performance of the surface absorber, because the coupled radiation itself is several orders of magnitude smaller than the propagating modes in the waveguide, and little will be reflected. The asymptotic convergence of the transition reflections in section~\ref{sec:absorber} and the $10^{-4}$ reflection coefficient in section~\ref{sec:introduction} consistently showed the excellent performance of surface absorbers, obviously unhampered by the effects of radiation.

\section{Conclusion}\label{conclusion}

We presented a novel numerical technique, a surface conductive absorber that is easily combined with the boundary element method, to eliminate the reflections due to the truncation of infinite channels. We illustrated this technique with a dielectric optical waveguide and described the modified BEM formulation needed to allow varying the surface conductivity. We further discussed the non-monotonically increasing decay rate with the surface conductivity and presented methods to calculate the decay rate using perturbation theory and Poynting's theorem. We demonstrated that the surface conductive absorber is orders of magnitude more effective than the volume conductive absorber, and showed an asymptotic power-law convergence of the transition reflection with respect to the length of the absorber to verify that the smoothness of conductivity function determines the transition reflection, also noted in \cite{oskooi08the}.

The major advantages of the surface conductive absorber are: (1) the varying surface conductivity is easily implemented in BEM and can significantly reduce the transition reflections; and (2) the volume properties of the absorber are the same as the waveguide, so there is no interior cross section to discretize, eliminating a potential source of numerical reflections.

In this paper, the surface conductive absorber is demonstrated with a simple example, a rectangular waveguide. Demonstrating that the method is as effective on more general structures is the subject of future work.

\section{Acknowledgement}
This work was supported by the Singapore-MIT Alliance (SMA), the MARCO Focus Center on Interconnect, in part by Dr. Dennis Healy of DARPA MTO, under award N00014-05-1-0700 administered by the Office of Naval Research, and the Army Research Office through the Institute for Soldier Nanotechnologies under contract DAAD-19-02-D0002.

\section{Appendix} \label{sec:appendix}
\subsection{Construction of a linear system}
From section~\ref{sec:formulation}, the five equivalent currents, $\mathbf{J}^L$, $\mathbf{M}^L$ on the waveguide surfaces, and $\mathbf{J}_e^R$, $\mathbf{J}_i^R$, $\mathbf{M}^R$ on the absorber surfaces, are to be determined by solving the equations (\ref{eq1}), (\ref{eq2}), (\ref{eq3}), (\ref{eq13}) and (\ref{eq14}). The currents are approximated with the RWG basis function \cite{rao82electromagnetic} on triangular-meshed surfaces,
\begin{eqnarray}
\mathbf{J} &=& \sum_m J_m\mathbf{X}_m(\mathbf{r}'), \label{RWGJ}\\
\mathbf{M} &=& \sum_m M_m\mathbf{X}_m(\mathbf{r}'), \label{RWGM}
\end{eqnarray}
where $\mathbf{X}_m(\mathbf{r}')$ is the RWG function on the $m$th triangle pair, and $J_m$, $M_m$ are the corresponding coefficients for the electric and magnetic currents, respectively. These unknown coefficients of the five equivalent currents assemble a vector $x$ to be solved for, specifically,
\begin{equation}
  \mathbf{x}=\left[
  \begin{array}{ccccc}
    J^L & M^L & J_e^R & J_i^R & M^R
  \end{array}
  \right]^T.
\end{equation}

Electric and magnetic fields are represented using the mixed-potential integral equation (MPIE)\cite{wang98a} for a low-order singularity, with integral operators $L$ and $K$ as in \cite{sheng98solution}
\setlength{\arraycolsep}{2pt}
\begin{eqnarray}
  \mathbf{E}_{l}^t(\mathbf{J},\mathbf{M}) &=& -Z_lL_{l}^t(\mathbf{J}) + K_{l}^t(\mathbf{M}),\label{eq15}\\
  \mathbf{H}_{l}^t(\mathbf{J},\mathbf{M}) &=& -K_{l}^t(\mathbf{J}) -
  \frac{1}{Z_l}L_{l}^t(\mathbf{M}),\label{eq16}
\end{eqnarray}
where $Z_l=\sqrt{\mu_l/\epsilon_l}$ is the intrinsic impedance, the subscript $l=e \mbox{ or } i$ denotes the exterior or interior region, and the superscript $t= L \mbox{ or } R$ denotes the waveguide or absorber surfaces. The integral operators on the $m$th RWG function are given by
\setlength\arraycolsep{0pt}{\small
\begin{eqnarray}
  L_{l}^t(\mathbf{X}_m) &=& jk_l\int_{S'}\left[ \mathbf{X}_m(\mathbf{r}') + \frac{1}{k_l^2}\nabla\nabla'\cdot\mathbf{X}_m(\mathbf{r}')
  \right]G_l(\mathbf{r},\mathbf{r}')dS' \nonumber\\\label{LOp}\\
  K_{l}^t(\mathbf{X}_m) &=& -\int_{S'} \nabla\times G_l(\mathbf{r},\mathbf{r}')\mathbf{X}_m(\mathbf{r}')
  dS',\label{KOp}
\end{eqnarray}}
where $S'$ is the surface of the $m$th triangle pair, $k_l=\omega\sqrt{\mu_l\epsilon_l}$ is the propagation constant in region $l$, and $G_l(\mathbf{r},\mathbf{r}')$ is the free-space Green's function in region $l$
\begin{equation}
  G_l(\mathbf{r},\mathbf{r}')=\frac{e^{-jk_l|\mathbf{r}-\mathbf{r}'|}}{4\pi|\mathbf{r}-\mathbf{r}'|},
\end{equation}
where $\mathbf{r}$ and $\mathbf{r}'$ are target and source positions, respectively.

We employ Galerkin method \cite{harrington68field} on the waveguide by using the RWG function as the testing function on target triangle pairs. The tested $L$, $K$ operators on the $n$th target triangle pair due to the $m$th source triangle pair become
\setlength\arraycolsep{5pt}
\begin{eqnarray}
\mathcal{L}_{l,nm}^t(\mathbf{X}_m)&=&\int_{S}\mathbf{X}_n(\mathbf{r})\cdot L_{l}^t(\mathbf{X}_m)dS,\label{testedLOp}\\
\mathcal{K}_{l,nm}^t(\mathbf{X}_m)&=&\int_{S}\mathbf{X}_n(\mathbf{r})\cdot K_{l}^t(\mathbf{X}_m)dS,\label{testedKOp}
\end{eqnarray}
where $S$ is the surface of the $n$th target triangle pair. Substituting the tested field operators into equations (\ref{eq1}), (\ref{eq2}) yields a matrix $\mathbf{A}_{LL}$ due to the currents $\mathbf{J}^L$, $\mathbf{M}^L$, and a matrix $\mathbf{A}_{LR}$ due to the currents $\mathbf{J}_e^R$, $\mathbf{J}_i^R$, $\mathbf{M}^R$.

On the absorber surfaces, the term $\sigma_e(\mathbf{r})\mathbf{E}^R_{l,\mathrm{tan}}$ in (\ref{eq13}) and (\ref{eq14}) requires another testing procedure in order to incorporate the surface conductivity
\begin{eqnarray}
\mathcal{L}_{l,nm}^{R\sigma_e}(\mathbf{X}_m)&=&\int_{S}\sigma_e(\mathbf{r})\mathbf{X}_n(\mathbf{r})\cdot L_{l}^R(\mathbf{X}_m)dS,\label{testedSigLOp}\\
\mathcal{K}_{l,nm}^{R\sigma_e}(\mathbf{X}_m)&=&\int_{S}\sigma_e(\mathbf{r})\mathbf{X}_n(\mathbf{r})\cdot K_{l}^R(\mathbf{X}_m)dS.\label{testedSigKOp}
\end{eqnarray}
Similarly, substituting the above four tested integral operators into (\ref{eq3}), (\ref{eq13}) and (\ref{eq14}) generate a matrix $\mathbf{A}_{RL}$ due to the currents $\mathbf{J}^L$, $\mathbf{M}^L$, and a matrix $\mathbf{A}_{RR}$ due to the currents $\mathbf{J}_e^R$, $\mathbf{J}_i^R$, $\mathbf{M}^R$.

Assembling the four matrices according to (\ref{eq1}), (\ref{eq2}), (\ref{eq3}), (\ref{eq13}) and (\ref{eq14}) yields a dense linear system
\begin{equation}
  \mathbf{A}\mathbf{x} = \mathbf{b}, \label{eq19}
\end{equation}
where
\begin{equation}
  \mathbf{A}=\left[
  \begin{array}{cc}
    \mathbf{A}_{LL} & \mathbf{A}_{LR}\\
    \mathbf{A}_{RL} & \mathbf{A}_{RR}
  \end{array}\right].
\end{equation}
The right-hand-side vector $\mathbf{b}$ contains tested incident electric and magnetic fields
\begin{equation}
  \mathbf{b}=\left[
  \begin{array}{ccccc}
    b_E^L & b_H^L & b_E^R & b_H^R & b_{E\sigma}^R
  \end{array}
  \right]^T,
\end{equation}
in which the $n$th entry of each subvector is given by
\begin{eqnarray}
b_{E,n}^t &=& \int_{S}\mathbf{X}_n(\mathbf{r})\cdot \mathbf{E}_{\mathrm{inc}}^t(\mathbf{r})dS,\\
b_{H,n}^t &=& \int_{S}\mathbf{X}_n(\mathbf{r})\cdot \mathbf{H}_{\mathrm{inc}}^t(\mathbf{r})dS,\\
b_{E\sigma, n}^R &=& \int_{S}\sigma_e(\mathbf{r})\mathbf{X}_n(\mathbf{r})\cdot \mathbf{E}_{\mathrm{inc}}^t(\mathbf{r})dS.
\end{eqnarray}

\subsection{Acceleration and Preconditioning with FFT }
\label{sec:acceleration}

The linear system (\ref{eq19}) can be solved with iterative algorithms, for instance, GMRES for this non-symmetrically dense system. In each iteration, the matrix-vector product takes $O(N^2)$ time, where $N$ is the number of unknowns. Moreover, to explicitly store the matrix $\mathbf{A}$ is expensive, requiring $O(N^2)$ memory. In fact, there have been many well-developed fast algorithms to reduce the costs of the integral equation solvers \cite{Phillips97a, Zhu03algorithms,zhang05rcs,song97multilevel}. In this paper, we use a straightforward and easily-implemented FFT-based fast algorithm to accelerate the SIE method on periodic guided structures.

\begin{figure}[!tb]
\begin{center}
\epsfclipon
\scalebox{0.55} {\epsffile{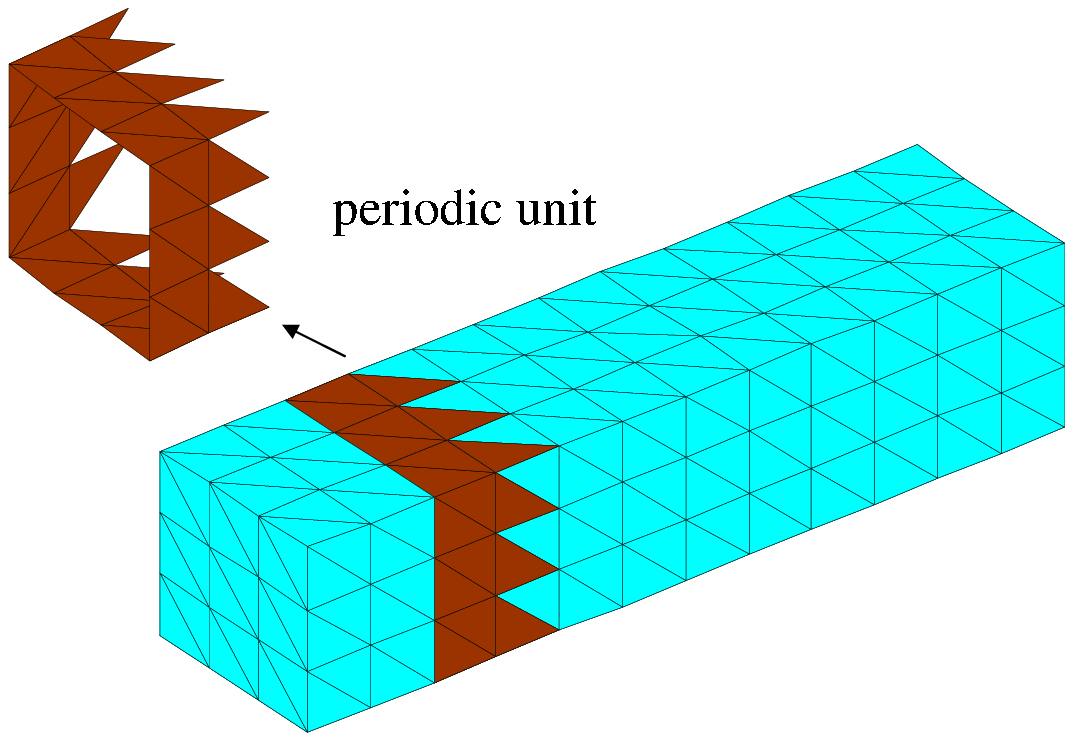}}
\caption{A discretized waveguide with a periodic unit.} \label{fig:period_WG}
\end{center}
\end{figure}

As shown in Fig.~\ref{fig:period_WG}, the waveguide is discretized into a periodically repeating set of the RWG triangle pairs. Due to the mesh periodicity and the space invariance of the operators (\ref{testedLOp}), (\ref{testedKOp}), the matrices $\mathbf{A}_{LL}$ and $\mathbf{A}_{LR}$ are block Toeplitz, requiring explicit calculation and storage of only a block row and a block column, reducing memory to approximately $O(N)$. A Toeplitz matrix can be embedded in a circulant matrix, and the circulant matrix-vector product can be computed with the FFT \cite{VanLoan92computational,Zhuang-96-acombined}. In this way, the computational costs are reduced approximately to $O(N \log N)$.

Because the surface conductivity $\sigma_e$ must vary with distance from the waveguide-absorber interface, the tested potential operators (\ref{testedSigLOp}), (\ref{testedSigKOp}) are not space invariant. Therefore, accelerating the $\sigma_e$ parameterized matrices $\mathbf{A}_{RL}$ and $\mathbf{A}_{RR}$ is not as straightforward as $\mathbf{A}_{LL}$ and $\mathbf{A}_{LR}$. Typically, the integral of (\ref{testedSigLOp}), (\ref{testedSigKOp}) is calculated numerically using Gauss quadrature \cite{trefethen97numerical}, summing up the tested potentials at quadrature (target) points with Gauss weights. The space invariance of the potential operators (\ref{LOp}), (\ref{KOp}) and periodicity of the mesh allows assembling a matrix of the potentials at target points by explicitly calculating only a block row and block column. Then the potentials at target points are summed after testing and multiplications with Gauss weights and surface conductivity, and eventually stamped into matrices $\mathbf{A}_{RL}$ and $\mathbf{A}_{RR}$.

Another great advantage of working with a Toeplitz or a block Toeplitz matrix is the existence of a highly efficient
preconditioner \cite{Strang86a, Chan88an,Chan90circulant}. A circulant matrix is approximated from the Toeplitz matrix, and then can be easily inverted with the FFT. We use this method to calculate a preconditioner for $\mathbf{A}_{LL}$, and use the block-diagonal preconditioner \cite{song97multilevel} for $\mathbf{A}_{RR}$.

 \vspace*{1mm}
 \scriptsize
\bibliographystyle{IEEEtran}
\bibliography{zhangbib}
\end{document}